\documentclass[aps,prb,twocolumn,groupedaddress,superscriptaddress]{revtex4-2}

\usepackage{graphicx}
\usepackage{dcolumn}
\usepackage{bm}
\usepackage[svgnames]{xcolor}
\usepackage{mathtools}
\usepackage{dsfont,amsfonts,bbold,amssymb}
\usepackage{url}
\usepackage{hyperref}
\usepackage{soul}
\usepackage[version=4]{mhchem}
\usepackage{braket}
\usepackage{tikz}

\begin{document}


\title{Tunable multi-magnon Floquet topological edge states}

\author{I. Martinez-Berumen}
\email{pablo.martinez2@mail.mcgill.ca}
\author{T. Pereg-Barnea}
\email{tamipb@physics.mcgill.ca}
\author{W. A. Coish}%
\email{coish@physics.mcgill.ca}

\affiliation{%
    Department of Physics, McGill University, Montr\'eal, Qu\'ebec, Canada H3A 2T8
}%

\date{\today}

\begin{abstract}
    We show that periodically time-modulating the Dzyaloshinskii-Moriya interaction (DMI) in a two-dimensional magnon insulator may induce a topological phase transition that results in the presence of robust edge modes. 
    To this end, we study a square lattice of spins interacting via an XXZ Heisenberg model with a ferromagnetic longitudinal coupling and antiferromagnetic transverse coupling, as well as the aforementioned time-modulated DMI. 
    The topologically protected edge states of this system are composed of coherent superpositions of single-magnon excitations and two-magnon bound states. 
    Furthermore, we show that the chirality of the edge states can be controlled by adjusting the relative phase for the drive on the DMI associated with nearest neighbors in the $x$ and $y$ directions.  
\end{abstract}

\maketitle

\section{Introduction} \label{sec:sec1}

The study of topological band structures has been extended from electronic systems~\cite{hasan2010colloquium,qi2011topological,chiu2016classification} to other areas such as photonics~\cite{lu2014topological,ozawa2019topological}, cold atoms~\cite{goldman2016topological,cooper2019topological}, and quantum magnetism~\cite{shindou2013topological,zhang2013topological,malki2020topological,wang2021topological,mcclarty2022topological,zhuo2023topological}. The topological magnon insulator (TMI) was proposed in 2013 by Zhang \textit{et al.}~\cite{zhang2013topological} motivated by the prediction of the magnon thermal Hall effect~\cite{katsura2010theory,matsumoto2011theoretical,matsumoto2011rotational} and its experimental observation in $\mathrm{Lu}_2\mathrm{V}_2\mathrm{O}_7$~\cite{onose2010observation,ideue2012effect}. TMIs are the magnonic analogues of Chern insulators, which host the quantum Hall effect~\cite{klitzing1980new,thouless1982quantized,haldane1988model}. Their band structure is often studied using linear spin-wave theory, where the Holstein-Primakoff transformation~\cite{holstein1940field} is applied to the spin Hamiltonian to obtain a bosonic system whose excitations (magnons) consist of a coherent superposition of single local spin-flip excitations out of the ferromagnetic ground state. In linear spin-wave theory, magnon-magnon interactions are neglected (in the low-temperature/dilute limit) and the Chern numbers of the free-magnon bands are used to distinguish between topologically trivial and topologically nontrivial phases.
The bulk-boundary correspondence predicts the existence of robust propagating edge states when the magnon insulator is in the nontrivial phase~\cite{mook2014edge}. This robustness makes TMIs interesting for potential applications in spin-wave devices like spin-wave diodes, splitters, and interferometers~\cite{wang2017topologically,wang2018topological}, and in potentially generating a strong coherent coupling between distant spins for quantum computing~\cite{hetenyi2022long}.

\begin{figure}
    \centering
    \includegraphics{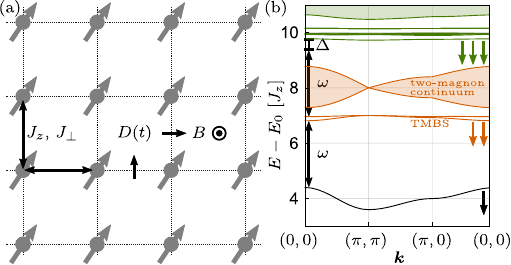}
    \caption{(a) Square lattice of spins with nearest-neighbor anisotropic spin-spin couplings ($J_z$, $J_\perp$), a time-modulated Dzyaloshinsky-Moriya interaction (DMI), $D(t)$, and a longitudinal Zeeman field $B$. (b) Energies for the single-magnon states (black $\downarrow$), two-magnon states (orange, above $\downarrow\downarrow$), and three-magnon states (green, above $\downarrow\downarrow\downarrow$) that are energies of the of the static Hamiltonian $H_\mathrm{S}$ [Eq.~\eqref{eq:H_S}]. The DMI is modulated with frequency $\omega$ so that it resonantly couples the single-magnon band to the two-magnon bound state (TMBS), but the modulation frequency is detuned by $\Delta$ from resonance with the three-magnon states. The spectrum is obtained by exact diagonalization (see Appendix~\ref{sec:exact_diagonalization}) in a $7\times 7$ lattice. The energy $E-E_0$ for each state is shown here relative to the ground state energy $E_0$ [Eq.~\eqref{eq:ground_state_energy}] using $J_\perp=J_z/5$, $B=2J_z$.}
    \label{fig:fig1}
\end{figure}

Magnon-magnon interactions come into play when at least two spin-flip excitations become probable. These interactions may alter the properties of a magnetic system, for example, by stabilizing a topological phase~\cite{mook2021interaction} or destroying it~\cite{habel2024breakdown}. For sufficiently strong attractive interactions, which are present in anisotropic quantum magnets, two local spin flips may bind together, forming a two-magnon bound state (TMBS) with lower energy than two separated spin flips~\cite{hanus1963bound,wortis1963bound,mook2023magnons}. The two-magnon spectrum is then divided into two components: The lower-energy part of the spectrum contains discrete TMBSs and the higher-energy part of the spectrum shows a two-magnon continuum, which may approximately describe two independent magnons [see Fig.~\ref{fig:fig1}(b)]. TMBSs have been observed experimentally in both ferromagnetic and antiferromagnetic systems~\cite{torrance1969excitation,petitgrand1980magnetic,fert1978excitation} and their dynamics have been simulated using cold atoms in optical lattices~\cite{fukuhara2013microscopic}. Furthermore, it has been proposed that the TMBS spectrum can be topologically nontrivial, resulting in robust edge states comprising a coherent superposition of local two-spin-flip pairs~\cite{qin2017topological,qin2018topological}. 

In systems where the total number of magnons is not conserved, it may be that magnonic excitations consist of hybridized single-magnon and two-magnon states~\cite{endoh1984resonant,heilmann1981one,bai2021hybridized,bai2023instabilities}. Mook \textit{et al.}~\cite{mook2023magnons} have shown that magnonic bands with this hybrid character can be topologically nontrivial in the presence of Dzyaloshinsky-Moriya interactions (DMI)~\cite{dzyaloshinskii1957thermodynamic,moriya1960anisotropic}. One of the systems considered in Ref.~\cite{mook2023magnons} is a square lattice of spins with an Ising anisotropy (an XXZ Heisenberg model) and DMI. This system can be mapped to a single-particle tight-binding problem on a Lieb lattice. The DMI is responsible for breaking an effective time-reversal symmetry (inducing an effective staggered flux in the Lieb lattice) and therefore allows the magnonic energy bands to be topologically nontrivial~\cite{mook2023magnons}. In this model, the presence of long-range (third-nearest-neighbor) interactions allows for the crossing of the single-magnon band with the TMBS. This crossing is needed to induce a topological phase transition. 
In this paper, we show that modulating the DMI in time is an alternative strategy to induce the band crossing and therefore to induce a topological phase transition, even in systems that may have very weak long-range spin couplings. 

Originally proposed in electronic systems~\cite{oka2009photovoltaic,lindner2011floquet,kitagawa2010topological}, the idea of using time-dependent fields to control the topological phase of a system (Floquet topological matter~\cite{rudner2020band,rudner2020floquet,oka2019floquet}) has also been applied to magnetic systems where the magnon number is conserved~\cite{owerre2017floquet,owerre2018photoinduced,zhang2021photoinduced,chen2025floquet}. In addition to being motivated by the development of Floquet topological matter, our work is also motivated by recent advances in the post-growth control of the strength of the DMI~\cite{kammerbauer2023dzyaloshinskii,meng2024regulation}. For example, it has been shown that the strength of the interfacial DMI can be modified and even change sign due to applied strain~\cite{gusev2020manipulation,cui2020strain,deger2020strain}, applied electric fields~\cite{yang2018controlling,zhang2018electrical,semenov2023electrical,liu2018analysis}, or by flipping the polarization of ferroelectric layers adjacent to the magnetic system~\cite{chen2024dzyaloshinskii}. These advances motivate us to investigate the possibility of inducing topologically nontrivial magnetic phases and to control them through a time-modulated DMI. 

The rest of this paper is organized as follows. In Sec.~\ref{sec:sec2}, we introduce the model, consisting of an XXZ Heisenberg Hamiltonian, a Zeeman field, and a periodically time-modulated DMI. In Sec.~\ref{sec:sec3}, we review the low-energy excitations of the XXZ Hamiltonian, the presence of TMBSs in the case of Ising anisotropy, and the presence of an effective time-reversal symmetry that results in a topologically trivial magnon insulator. We also review the idea of hybrid magnons, i.e. eigenstates comprising a superposition of single-magnon and TMBSs hybridized by the DMI. In Sec.~\ref{sec:sec4}, we analyze the system with the time-modulated DMI. We obtain an effective time-independent Hamiltonian and study the topological properties (Chern numbers for the low-energy bands) and the associated conditions for the presence of robust edge modes. We show that in the case of a static DMI there are no edge modes present due to the absence of an energy gap and that this can be remediated by driving the DMI. Additionally, we show that the chirality of the edge states can be tuned by controlling a phase introduced between the time-modulated DMI acting between neighboring spins in the $x$ and $y$ directions (Sec.~\ref{sec:sec4C}) and discuss the potential realization of our model (Sec.~\ref{sec:sec4D}). Finally, in Sec.~\ref{sec:sec5}, we summarize our results.

\section{Model} \label{sec:sec2}

We consider a two-dimensional square lattice of spins [Fig.~\ref{fig:fig1}(a)] described by the time-dependent Hamiltonian
\begin{equation} \label{eq:Ht}
    H(t) = H_\mathrm{S} + H_\mathrm{DM}(t),
\end{equation}
where $H_\mathrm{DM}(t)$ describes a time-dependent DMI and where $H_\mathrm{S}$ is a (static) XXZ Hamiltonian with a Zeeman field in the $z$ direction (setting $\hbar=1$):
\begin{equation}  \label{eq:H_S}
    \begin{split}
    H_\mathrm{S} = \sum_{\braket{\bm r,\bm r'}}&\left[-J_z S^z_{\bm r} S^z_{\bm r'} + J_\perp \left(S_{\bm r}^x S^x_{\bm r'} + S_{\bm r}^y S^y_{\bm r'} \right) \right] \\&- B \sum_{\bm r}S^z_{\bm r}.
\end{split}
\end{equation}
Here, $\bm S_{\bm r}= (S^x_{\bm r},S^y_{\bm r},S^z_{\bm r})$ is the spin-1/2 operator associated with site $\bm r$, $J_z>0$ is a ferromagnetic longitudinal coupling, and $J_\perp>0$ is an antiferromagnetic transverse coupling. The signs of $J_z$ and $J_\perp$ will be important in what follows. The parameter $B$ is the effective Zeeman field. The lattice consists of $N=L\times L$ sites so that the sum runs over $\bm r=n_x\bm e_x+n_y\bm e_y$, where $n_x,n_y=1,\hdots,L$ (setting the lattice constant $a=1$), and where $\bm e_i$ is the unit vector in the $i$ direction. 

The time-modulated DMI is assumed to take the form:
\begin{equation} \label{eq:H_DM}
\begin{split}
    H_\mathrm{DM}(t) = \sum_{\bm r} \big[ D_x(t)& \,(\bm S_{\bm r} \times \bm S_{\bm r+\bm e_x} )_y \\ &+ D_y(t) \,   (\bm S_{\bm r} \times \bm S_{\bm r+\bm e_y} )_x \big].
\end{split}
\end{equation}
We assume that the DMI can be modulated in time via, e.g., a symmetry-breaking time-dependent electric field or by inducing a time-dependent strain. In either case, it may be possible to independently control the strength of the DMI associated with nearest neighbors along $x$ and $y$ [$D_x(t)\neq D_y(t)$]. We will show in Sec.~\ref{sec:sec4C}, below, that this independent control can be used to change the chirality of the resulting topologically protected edge states. However, for most of this paper, we will restrict to the case where the two coupling parameters are equal and in-phase: 
\begin{equation} \label{eq:Dt}
    D_x(t)=D_y(t)=D(t) = D_0 + 2d \cos \omega t.
\end{equation}
The DMI strength, $D(t)$, oscillates with amplitude $2d$ and frequency $\omega$ around the average value $D_0$, making the Hamiltonian periodic $H(t+T)=H(t)$ with period $T=2\pi/\omega$. 

We work in the regime given by
\begin{equation} \label{eq:parameter_regime}
    J_z \gg J_\perp, D_0, \quad J_\perp \gg d.
\end{equation}
The inequality $J_z\gg J_\perp,D_0$ guarantees a ferromagnetic ground state and allows us to treat both the transverse terms and the DMI as a perturbation to the Ising Hamiltonian. Additionally, $J_z\gg J_\perp$ guarantees the presence of TMBSs as the energetic advantage of having two adjacent spin flips, $J_z$, separates these states from the two-magnon continuum, whose bandwidth scales with $J_\perp$ [Fig.~\ref{fig:fig1}(b)]. The condition $J_\perp\gg d$ guarantees a weak off-resonant coupling between the TMBSs and the three-magnon states when driving the DMI; the detuning from resonance $\Delta$ [Fig.~\ref{fig:fig1}(b)] is of order $J_\perp$ (see Sec.~\ref{sec:sec4}, below), while the coupling between the TMBSs and the three-magnon subspace is of order $d$.

Before continuing with the analytical treatment, we comment on the choice of an antiferromagnetic transverse coupling. For an antiferromagnetic coupling, the maximum of the single-magnon band coincides with the minimum of the TMBS bands at $\bm k =(0,0)$ [Fig.~\ref{fig:fig1}(b)]. This admits a band inversion (via resonant driving), resulting in two Floquet bands with an energy gap that allows for the presence of robust edge states. For a ferromagnetic transverse coupling, the band minima of the magnon and TMBS bands would lie at the same point in the Brillouin zone. In this case, driving the DMI to induce the band crossing would not be a viable strategy in the regime given by Eq.~\eqref{eq:parameter_regime}, because the bands would overlap in energy whenever a level crossing could be induced. 

In the next section, we review the ground state and low-energy excitations of the static Hamiltonian $H_\mathrm{S}$ and the effects of a static DMI [$D_0\neq 0$, $d=0$ in Eq.~\eqref{eq:H_DM}]. Then, in Sec.~\ref{sec:sec4} we show how topologically protected Floquet edge states can be engineered by driving the DMI according to Eq.~\eqref{eq:Dt}. 

\section{Hybrid magnons} \label{sec:sec3}

In this section we first review the single-magnon and two-magnon excitations of the XXZ Heisenberg model [Eq.~\eqref{eq:H_S}] and define the notation used in the rest of the paper (Secs.~\ref{sec:sec3A} and~\ref{sec:sec3B}). Then, in Sec.~\ref{sec:sec3C}, we follow Ref.~\cite{mook2023magnons} and review how the single-magnon states are hybridized with the TMBSs via the DMI and how the effective time-reversal symmetry is broken, allowing these hybridized bands to be topologically nontrivial. Finally, we show that for a static DMI [$D_0\neq 0$, $d=0$ in Eq.~\eqref{eq:H_DM}], the nontrivial bands overlap, hindering the presence of robust topologically-protected edge states. 

\subsection{Magnons} \label{sec:sec3A}
We first consider the ground state and low-energy excitations of the static Hamiltonian $H_\mathrm{S}$. For a large Ising anisotropy, $J_z\gg J_\perp$, the XXZ Hamiltonian has a ferromagnetic ground state with every spin ``up'' ($S^z_{\bm r}=+1/2$) when $B>0$:
\begin{equation}
    H_\mathrm{S}\ket{0}=E_0\ket{0},\quad \ket{0} = \ket{\uparrow,\hdots,\uparrow},
\end{equation}
with ground-state energy
\begin{equation} \label{eq:ground_state_energy}
     E_0 = -\frac{1}{2}N\left(B+J_z\right).
\end{equation}

Since $H_\mathrm{S}$ commutes with the $z$-component of total spin, $[H_\mathrm{S},\sum_{\bm r} S_{\bm r}^z]=0$, it can be diagonalized in sectors having a definite number of spin flips. The lowest-energy excitations, composed of a linear combination of single spin flips, are the spin-wave (magnon) states $\ket{\bm k}$:
\begin{equation}
   \ket{\bm k} = \frac{1}{\sqrt{N}} \sum_{\bm r} e^{i\bm k\cdot \bm r}\ket{\bm r}, \quad \ket{\bm r} \equiv S_{\bm r}^- \ket{0},
\end{equation}
where $\ket{\bm r}$ is the state with a spin flip at site $\bm r$ and $S^\pm_{\bm r} = S^x_{\bm r} \pm iS^y_{\bm r}$ are the spin raising/lowering operators. The excitation energy (relative to $E_0$) of a magnon with wavevector $\bm k$ is given by
\begin{equation} \label{eq:E_k}
    E_{\bm k} = B + 2J_z + J_\perp\left(\cos k_x + \cos k_y \right).
\end{equation}
Figure~\ref{fig:fig1}(b) shows the dispersion relation for the single-magnon states when $J_\perp=J_z/5$ and $B=2J_z$. Since the static Hamiltonian conserves the number of spin flips, the single-magnon band is decoupled from the rest of the Hilbert space and it is always topologically trivial (a single band cannot have a nonzero Chern number). 

\begin{figure}
    \centering
    \includegraphics[]{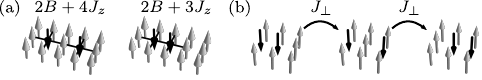}
    \caption{(a) States with two spin flips and their associated energies in the Ising limit $J_\perp=0$. (b) Hopping of a spin flip produced by the transverse exchange interactions in states with two spin flips.}
    \label{fig:fig2}
\end{figure}

\subsection{Two-magnon bound states} \label{sec:sec3B}

After the single-magnon states, the states that are next highest in energy lie in the subspace of states with two spin flips, 
\begin{equation} \label{eq:two_spin_flips}
    \ket{\bm r,\bm r'} \equiv S^-_{\bm r}S_{\bm r'}^-\ket{0}.
\end{equation}
To obtain the eigenstates of $H_\mathrm{S}$ in this subspace, we first consider the Ising limit ($J_\perp=0$) and then treat the transverse coupling (term proportional to $J_\perp$) as a perturbation. In the Ising limit, the states $\ket{\bm r,\bm r'}$ are eigenstates with excitation energy $2B + 3J_z$ for adjacent spin flips and $2B + 4J_z$ otherwise. The energetic cost associated with non-adjacent spin flips ($J_z$) arises because adjacent spin flips have only six antiparallel nearest neighbors (instead of eight) [Fig.~\ref{fig:fig2}(a)]. The subspace with two spin flips is then separated into a low-energy sector composed of two-magnon bound states (TMBSs) ($\ket{\bm r,\bm r + \bm e_x}$ and $\ket{\bm r,\bm r + \bm e_y}$), and a high-energy sector composed of the states $\ket{\bm r,\bm r'}$ with $|\bm r-\bm r'|>1$. 

The transverse coupling induces hopping for spin flips and it couples the low-energy and high-energy sectors within the space of two spin flips, as it separates spin flips that are initially adjacent [Fig.~\ref{fig:fig2}(b)]. An effective Hamiltonian for the low-energy subspace can be obtained using second-order perturbation theory. The Hamiltonian contains the effective coupling between two low-energy states via a (higher-energy) virtual state as depicted in Fig.~\ref{fig:fig2}(b), where two actions of the perturbation results in an effective hopping from the state $\ket{\bm r,\bm r + \bm e_x}$ to the state $\ket{\bm r + \bm e_x,\bm r+ 2\bm e_x}$. 

Before writing down the effective Hamiltonian, we note that the Hamiltonian $H(t)$ has translational symmetry. Hence, it is convenient to Fourier transform the two-spin-flip states [Eq.~\eqref{eq:two_spin_flips}] with respect to the center point between the two spin-flip sites:
\begin{equation} \label{eq:tm_def}
    \ket{\bm k,\bm r} = \frac{1}{\sqrt{N}} \sum_{\bm r'} e^{i\bm k\cdot (\bm r'+\bm r/2)} \ket{\bm r',\bm r' + \bm r}.
\end{equation}
The states above are not eigenstates of $H_\mathrm{S}$, but the Hamiltonian preserves the wavevector $\bm k$: $\bra{\bm k,\bm r}H_\mathrm{S}\ket{\bm k',\bm r'} \propto \delta_{\bm k,\bm k'}$. This allows us to associate a wavevector to the two-magnon states and therefore to study their topological properties in terms of a Chern number, derived from effective Bloch states defined over the first Brillouin zone~\cite{qin2018topological}.

The TMBSs can be written in terms of the Fourier transforms [Eq.~\eqref{eq:tm_def}] as 
\begin{align} \label{eq:tmbs_x}
    \ket{\bm k,\bm e_x} \equiv \frac{1}{\sqrt{N}} \sum_{\bm r} e^{i \bm k \cdot (\bm r + \bm e_x/2)} \ket{\bm r,\bm r+ \bm e_x}, \\ 
    \label{eq:tmbs_y}
    \ket{\bm k,\bm e_y} \equiv \frac{1}{\sqrt{N}} \sum_{\bm r} e^{i \bm k \cdot (\bm r + \bm e_y/2)} \ket{\bm r,\bm r+ \bm e_y}.
\end{align}
These states are coupled to the two-magnon continuum via the transverse coupling in $H_\mathrm{S}$. When there is a large Ising anisotropy $J_z\gg J_\perp$, an effective Hamiltonian for the low-energy sector can be obtained using second-order perturbation theory (see Appendix~\ref{sec:Appendix_effective_Hamiltonian}). Each block of the effective low-energy Hamiltonian, labeled by wavevector $\bm k$, is given by 
\begin{equation} \label{eq:H_eff_1}
    h_{\bm k} = \left(2B+3J_z\right)\sigma_0 - \frac{J_\perp^2}{J_z}\left(\begin{array}{cc}
       c^2_x+2c^2_y  &  2c_xc_y \\
        2c_xc_y & 2c^2_x+c^2_y 
    \end{array}\right),
\end{equation}
in the basis  $(\ket{\bm k,\bm e_x},\,\, \ket{\bm k,\bm e_y})^T$ where $\sigma_0$ is the $2\times 2$ identity matrix, and we have used the abbreviation $c_i = \cos(k_i/2)$. 

The Hamiltonian $h_{\bm k}$ is gapless at $\bm k=(\pi,\pi)$ and therefore it cannot have well defined Chern numbers, so it cannot host topologically protected edge states. More fundamentally, the Hamiltonian matrix elements $\bra{\bm k,\bm r}H_\mathrm{S}\ket{\bm k,\bm r'}$ are purely real [see Eq.~\eqref{eq:H_S_on_kr} in Appendix \ref{sec:exact_diagonalization}], which means that $H_\mathrm{S}$ has an effective time-reversal symmetry (TRS), which prohibits the bands from having a non-zero Chern number. Although TRS is already broken in a system with magnetic ordering, it is the effective (and not the real) TRS that must be broken in order to have non-zero Chern numbers. A Hamiltonian with translational symmetry described by the matrix $\mathcal{H}_{\bm k}$ has an effective TRS when $U^\dagger \mathcal H_{\bm k'} U = \mathcal H_{\bm k}^*$ is satisfied for a unitary transformation $U$ and an orthogonal transformation of the momentum: $\bm k'=O\bm k$. In the presence of this symmetry, the Berry curvature has opposite signs at wavevectors $\bm k,\bm k'$ and therefore the integrated Berry curvature over the BZ vanishes (see Appendix~\ref{sec:Appendix_time_reversal}). Since a real Hamiltonian satisfies $\mathcal H_{\bm k}^*=\mathcal H_{\bm k}$, it always has an effective TRS (in this case both $U$ and $O$ correspond to the identity operator). 

In the next section, we show that although a static DMI ($D_0\neq 0$, $d=0$) breaks the effective TRS, its mere presence is not enough to obtain topologically protected edge states, as the absence of an energy gap may hinder their existence. However, we show that by driving the DMI ($d\neq 0$) we can obtain a (quasi-)energy gap that guarantees the presence of topologically protected edge states (Sec.~\ref{sec:sec4}).

\subsection{Hybridized magnon bands} \label{sec:sec3C}

The DMI breaks rotational symmetry about the $z$-direction and it couples states having a number of spin flips that differs by one. The ferromagnetic ground state $\ket{0}$, remains an eigenstate, even in the presence of DMI as $H_\mathrm{DM}(t)\ket{0}=0$. In momentum space, the coupling between the single-magnon state and the TMBSs is  given by
\begin{align} 
    \braket{\bm k|H_\mathrm{DM}(t)|\bm k,\bm e_x}&=-iD(t) s_x, \\
    \braket{\bm k|H_\mathrm{DM}(t)|\bm k,\bm e_y}&=D(t)s_y,
\end{align}
where $s_i=\sin(k_i/2)$. The DMI acts locally, coupling the single-magnon state only to the TMBSs (and not to the two-magnon continuum). The DMI also preserves translation symmetry and is therefore diagonal in momentum space. 

\begin{figure}
    \centering
    \includegraphics{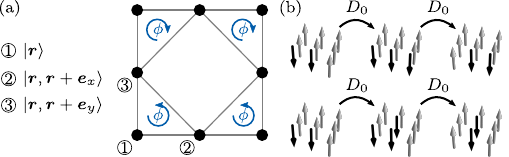}
    \caption{(a) Effective Lieb lattice describing the low-energy subspace composed of single- and two-adjacent spin flips. The arrows indicate the direction in which the phase $\phi=\pi/2$ is acquired. (b) Some second-order processes contributing to matrix elements in Eq.~\eqref{eq:H_eff_2}. The virtual three-magnon states mediate an effective coupling between the TMBSs.}
    \label{fig:fig3}
\end{figure}

In addition to coupling the single-magnon state and the TMBSs, the DMI breaks the effective TRS~\cite{mook2023magnons}. The breaking of the effective time-reversal symmetry can be understood by considering the low-energy subspace spanned by the single-magnon state and the TMBSs, in real space: \{$\ket{\bm r}$, $\ket{\bm r,\bm r+\bm e_x}$, $\ket{\bm r,\bm r+\bm e_y}$\}. These states can be associated with the position of a particle hopping on a Lieb lattice, where the single spin flips correspond to the particle located at position $\bm r$ and adjacent spin flips describe a particle located at the midpoint between the two spin flips, at $\bm r + \bm e_i/2$, as illustrated in Fig.~\ref{fig:fig3}(a). Without the DMI, the single-spin-flip states are decoupled from the two-adjacent spin-flip states. The DMI couples these states inducing an effective staggered flux, similar to the case of the Haldane model~\cite{haldane1988model}. Here, as in the Haldane model, there is a nonzero local flux within part of a unit cell, but there is a net zero flux through the complete unit cell. The presence of this flux that cannot be gauged away can be understood from the matrix elements coupling the single spin flips and two-adjacent spin flips in the $y$-direction: $\bra{\bm r,\bm r \pm  \bm e_y}H_\mathrm{DM}(t)\ket{\bm r}=\pm i D(t)/2$. Other hopping terms are real and don't contribute to the flux. The mapping of this problem to the Lieb lattice and the relationship to an effective time-reversal symmetry is also discussed at length in Ref.~\cite{mook2023magnons}. 

Before moving to the driven case, here we review the case, considered in Ref.~\cite{mook2023magnons}, of a static DMI ($D_0 \neq 0$, $d=0$) in the limit of large Ising anisotropy and weak DMI: $J_z \gg J_\perp,D_0$. Applying second-order perturbation theory in both the transverse exchange $J_\perp$ and the DMI $D_0$ (Appendix~\ref{sec:app_Eff_H_hs}), the resultant low-energy Hamiltonian, now written in the basis $(\ket{\bm k}, \,\,\ket{\bm k,\bm e_x},\,\, \ket{\bm k,\bm e_y})^T$, is given by 
\begin{equation} \label{eq:H_eff_2}
    H_{\bm k} =
    \left(\begin{matrix}
       E_{\bm k} & -i D_0 s_x & D_0 s_y \\
       i D_0 s_x & E_{\bm k}^x & g_{\bm k} \\ 
       D_0 s_y & g_{\bm k}^* & E_{\bm k}^y \\
    \end{matrix}\right),
\end{equation}
where
\begin{align} \label{eq:epsilon_x} 
    E_{\bm k}^{x} &= 2B + 3J_z - \frac{J_\perp^2\left(c_{x}^2+2c_{y}^2\right)}{J_z}-\frac{D_0^2\left(2-c_{x}^2\right)}{B+J_z}, \\ \label{eq:epsilon_y}
    E_{\bm k}^{y} &= 2B + 3J_z - \frac{J_\perp^2\left(2c_{x}^2+c_{y}^2\right)}{J_z}-\frac{D_0^2\left(2-c_{y}^2\right)}{B+J_z},\\  \label{eq:gk}
    g_{\bm k} &= - 2 \frac{J_\perp^2}{J_z} c_x c_y - \frac{i}{2} \frac{D_0^2}{B+J_z} s_x s_y.
\end{align}
In addition to the coupling between the single magnon and the TMBSs, the TMBSs are coupled to the three-magnon states via the DMI. The three-magnon states mediate an effective second-order coupling between the TMBSs [see Fig.~\ref{fig:fig3}(b)].

Figure~\ref{fig:fig4}(a) shows the spectrum of Eq.~\eqref{eq:H_eff_2} for $J_\perp=J_z/5$, $D_0=J_z/15$, and $B=2J_z$. The DMI produces a band opening at $\bm k=(\pi,\pi)$ and therefore the Chern numbers of the three bands are well defined and are given by $(0,1,-1)$ from the lowest to the highest band. The Chern numbers are computed using the method by Fukui \textit{et al}.~\cite{fukui2005chern} for a discrete Brillouin zone. To apply this method properly, the Hamiltonian $H_{\bm k}$ must be periodic in momentum space. To accomplish this, we apply the unitary gauge transformation given by $U_{\bm k}=\mathrm{diag}(1,e^{-ik_y/2},e^{-ik_x/2})$.

The number of edge modes traversing an energy gap is given by the sum of the Chern numbers of the bands below this gap~\cite{mook2014edge}. Since the lowest band has a zero Chern number, no edge modes are expected to traverse the energy gap between the lowest and the middle band. We would expect edge modes between the middle and highest bands if these were separated by an energy gap. This is not the case in the regime given by $J_\perp\gg D_0$. The lowest energy of the upper band is $2B+3J_z-J_\perp^2/J_z$ for $\bm k=(0,0)$ and is below the maximum energy of the middle band, given approximately by $\approx 2B+3J_z-D^2_0/(B+J_z)$ at $\bm k = (\pi,\pi)$. The presence of topological edge states is therefore not guaranteed. This situation can be remediated by the presence of sufficiently strong next-nearest-neighbor interactions~\cite{mook2023magnons}. Here, however, we propose that in systems where these long-range interactions are too weak, time-modulating the DMI is an alternative strategy to produce edge modes. 

\section{Floquet topological edge states} \label{sec:sec4}

\begin{figure}
    \centering
    \includegraphics[]{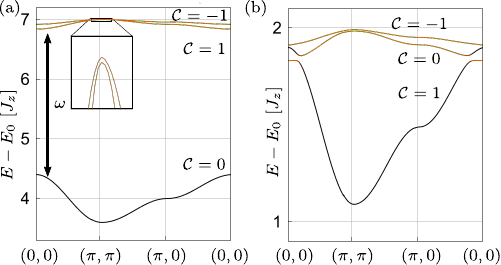}
    \caption{(a) Bulk band structure with static DMI. In the undriven system, the Chern numbers are $(0,1,-1)$. Driving the DMI with frequency $\omega$ resonantly couples the single-magnon band to the TMBS bands. (b) Bulk quasienergies of the Hamiltonian given by Eq.~\eqref{eq:H_eff_3} (in the rotating frame). The color represents single-magnon (black) and TMBS character (orange/gray). The quasienergies for the system with open boundary conditions and the corresponding edge states are shown in Fig.~\ref{fig:fig6}.}
    \label{fig:fig4}
\end{figure}

For weak DMI, $J_\perp \gg D_0$, the presence of robust edge modes is not guaranteed due to the absence of an energy gap between the topologically nontrivial bands. In this section, we show that topological magnon bands with an energy gap can be induced by driving the DMI. The goal is to produce a band inversion between the single-magnon band and the lower TMBS band by driving the DMI with a frequency that compensates for this gap [Fig.~\ref{fig:fig4}(a)].  

\subsection{Floquet bands} \label{sec:sec4A}

The topological properties of a periodic Hamiltonian with period $T$: $H(t+T)=H(t)$ can be studied in terms of its Floquet states. The Floquet states $\{\ket{\psi_\alpha(t)}\}$ form a set of orthogonal solutions to the time-dependent Schr\"odinger equation $i\frac{d}{dt}\ket{\psi_\alpha(t)} = H(t)\ket{\psi_\alpha(t)}$ (setting $\hbar=1$). These solutions have the form
$\ket{\psi_\alpha(t)} = e^{-i\varepsilon_\alpha t} \ket{\phi_\alpha(t)}$, where $\ket{\phi_\alpha(t)}$ has the same periodicity as the Hamiltonian:  $\ket{\phi_\alpha(t+T)}=\ket{\phi_\alpha(t)}$ and $\varepsilon_\alpha$ is known as the quasienergy~\cite{rudner2020floquet,rudner2020band}. At stroboscopic times $t=nT$, the Floquet state is periodic up to a phase that is proportional to the quasienergy, $\ket{\psi_\alpha(nT)}=e^{-i\varepsilon_\alpha nT}\ket{\psi_\alpha(0)}$, as would be true for an eigenstate evolving under a time-independent Hamiltonian with energy $\varepsilon_\alpha$. The state $\ket{\psi_\alpha(0)}$ is then the eigenstate of a time-independent Hamiltonian, the Floquet Hamiltonian $H_\mathrm{F}$, defined by the time-evolution operator over one period:
\begin{equation}
    U(T,0)=\mathcal T e^{-i\int_0^T dt'H(t')} \equiv e^{-iH_\mathrm{F} T},
\end{equation}
where $\mathcal{T}$ is the time-ordering operator~\cite{rudner2020floquet,rudner2020band}. The topological properties of the time-dependent Hamiltonian $H(t)$ can then be studied using the Floquet states $\ket{\psi_\alpha(0)}$.

To obtain the Floquet states we go into a rotating frame given by $\ket{\psi(t)} = U(t) \ket{\tilde \psi(t)}$ where $U(t)=\exp\left(-i\omega S_\mathrm{sf}t\right)$, where $S_\mathrm{sf}=N/2-\sum_{\bm r}S_{\bm r}^z$ counts the number of spin flips. The Hamiltonian in the rotating frame is given by 
\begin{equation} \label{eq:H_rotating}
\begin{split}
    \tilde H(t) &= U^\dagger(t) H(t) U(t) - iU^\dagger(t) \dot U(t)  \\&= U^\dagger(t) H(t) U(t) - \omega S_\mathrm{sf} \\&= H_\mathrm{S}+U^\dagger(t)H_\mathrm{DM}(t)U(t)-\omega S_\mathrm{sf}.
\end{split}
\end{equation}
The static Hamiltonian $H_\mathrm{S}$ commutes with $S_\mathrm{sf}$ (as it preserves the number of spin flips) and hence it is unaffected by the unitary transformation. The last term in Eq.~\eqref{eq:H_rotating} shifts the $n$-magnon band down in energy by $n\omega$, reducing the spacing between pairs of bands associated with a different number of magnons. The second term in Eq.~\eqref{eq:H_rotating} includes a time-independent DMI with amplitude $d$, and fast oscillating DMI terms [with amplitudes $D_0$ and $d$, cf.~\eqref{eq:rotating_dmi} in Appendix~\ref{sec:app_Eff_H_ht}]. These terms can be neglected using the rotating-wave approximation (RWA) in the regime given by $d,D_0\ll J_z$. Then, in the rotating frame the Hamiltonian can be approximated by
\begin{equation}
\begin{split}
    \tilde H \approx &H_\mathrm{S} - \omega S_\mathrm{sf} \\
    &+d \sum_{\bm r} \big[ (\bm S_{\bm r} \times \bm S_{\bm r+\bm e_x} )_y + \bm  (\bm S_{\bm r} \times \bm S_{\bm r+\bm e_y} )_x \big].
\end{split}
\end{equation}
Here, as in Sec.~\ref{sec:sec3C}, we apply second-order perturbation theory in the transverse exchange $J_\perp$ and the DMI (now with amplitude $d$), valid in the regime $d\ll J_\perp$, $J_\perp \ll J_z$, and obtain the effective Hamiltonian in momentum space
\begin{equation} \label{eq:H_eff_3}
    \tilde H_{\bm k} = 
    \left(\begin{matrix}
       E_{\bm k}-\omega & -i d s_x & d s_y \\
       i d s_x & \tilde E_{\bm k}^x-2\omega &  \tilde g_{\bm k} \\ 
       d s_y & \tilde g_{\bm k}^* & \tilde E_{\bm k}^y-2\omega \\
    \end{matrix}\right),
\end{equation}
where 
\begin{align}
    \tilde E_{\bm k}^x &= 2B + 3J_z  -\frac{J_\perp^2\left( c_x^2+2c_y^2 \right)}{J_z} - \frac{d^2\left(2-c_x^2\right)}{B+J_z-\omega}, \\
    \tilde E_{\bm k}^y &= 2B + 3J_z - \frac{J_\perp^2\left( c_y^2+2c_x^2 \right)}{J_z} - \frac{d^2\left(2-c_y^2\right)}{B+J_z-\omega} , \\
    \tilde g_{\bm k} &= -\frac{2J_\perp^2}{J_z}c_x c_y - i \frac{d^2}{B+J_z-\omega} s_xs_y.
\end{align}
Note that, since the transformation to the rotating frame through $U(t)$ preserves the number of spin-flip excitations, the Floquet states remain in a superposition of states with different magnon number even when the Floquet states are nonstationary when rewritten in the lab frame.

After comparing with the effective Hamiltonian for the case of static DMI, Eq.~\eqref{eq:H_eff_2}, we observe two main differences. First, the spacing between the single-magnon band and the TMBS band is reduced by $\omega$. Second, there is a modification of the second-order corrections (terms proportional to $d^2$) coming from coupling to the three-magnon states. This is because the three-magnon states are also shifted by an amount $\omega$ closer to the TMBS bands. 

A topological phase transition might occur if the gap between the single-magnon band and the TMBS bands closes. This occurs when $\omega$ is greater than the energy gap:
\begin{equation} \label{eq:lower_bound}
    \omega > B + J_z - 2J_\perp - \frac{5J_\perp^2}{J_z}.
\end{equation}
The band gap of the undriven system $H_\mathrm{S}$, given by the RHS of the inequality above, is the energy difference between the lowest energy of the TMBS bands and the highest point of the single-magnon band, both located at $\bm k=(0,0)$ and given by $2B+3J_z-5J_\perp^2/J_z$ and $B+2J_z+2J_\perp$ (up to second-order corrections in $J_\perp$), respectively. In this range of frequencies, the single-magnon band is shifted up in energy, crossing the lowest TMBS band and the bands undergo a topological phase transition. Figure~\ref{fig:fig4}(b) shows an example for $J_\perp=J_z/5$, $B=2J_z$, $D=J_z/15$, $\lambda = 1$, and $\omega=2.5 J_z$. The Chern number of the lowest band goes from $0$ to $1$ and the Chern number of the middle band goes from $1$ to $0$. The bands have both single-magnon and TMBS character (see color scale). Contrary to the static case, there is an energy gap between the lower band and the middle band, allowing for the presence of robust edge modes. 

To induce a band inversion that guarantees the presence of topologically protected edge states, the driving frequency $\omega$ must be larger than the lower bound [Eq.~\eqref{eq:lower_bound}]. As we increase the frequency, the TMBSs will eventually become resonantly coupled to the three-magnon states. This sets an upper bound for $\omega$ as our perturbative approach is valid only if the detuning $\Delta$ between the TMBSs and the three-magnon states is large compared to the amplitude of the oscillating DMI $d$. This upper bound is the energy difference between the lowest three-magnon state, given by $3B+4J_z-J_\perp$ (Appendix~\ref{sec:3mbs_perturbation}), and the highest energy of the TMBSs, given by $2B+3J_z$. The detuning is
\begin{equation} \label{eq:Delta}
    \Delta = B+J_z - J_\perp -\omega.
\end{equation}
Our perturbative approach is then valid when $\Delta \gg d$. The detuning must be positive to avoid coupling the TMBSs to higher-energy three-magnon states. This means that 
\begin{equation}
    \omega < B+J_z-J_\perp    
\end{equation}
must be satisfied. This condition together with Eq.~\eqref{eq:lower_bound} result in
\begin{equation} \label{eq:inequalities_Delta}
    0<J_\perp - \Delta < J_\perp.
\end{equation}
Therefore, the detuning $\Delta$ is at most of order $J_\perp$. Aiming to maximize the quasi-energy gap, which is of order $d$, means driving with a frequency that is close to this maximal value for the detuning, i.e. $\Delta \sim J_\perp$. In this regime, our perturbative approach is valid when $J_\perp \gg d$. 

\begin{figure}
    \centering
    \includegraphics{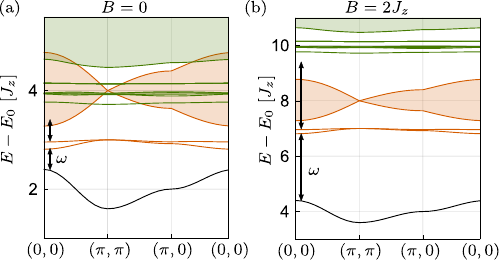}
    \caption{Exact diagonalization of the static Hamiltonian $H_\mathrm{S}$ with $J_\perp = J_z/5$ and two values of the magnetic field $B$. (a) In the absence of a magnetic field ($B=0$), the frequency $\omega$ required to resonantly couple the single-magnon and TMBS bands also resonantly couples the TMBSs to the two-magnon continuum. (b) A magnetic field $B$ can be used to avoid these undesired resonances.}
    \label{fig:fig5}
\end{figure}

The magnetic field $B$ does not play a role in the equations describing the regime of validity of our perturbative analysis and is not necessary for the physics discussed above. This is because the magnetic field increases the energy separation between magnon bands with different magnon number [Fig.~\ref{fig:fig5}]. To produce the resonant coupling between the single-magnon and the TMBS bands, this energy difference must be overcome [Eq.~\eqref{eq:lower_bound}]. This is explicit in the diagonal matrix elements of Eq.~\eqref{eq:H_eff_3}. Every term containing $B$ has at least one $-\omega$ term. Nevertheless, the magnetic field can be used as a handle to avoid resonances between the TMBSs and the two-magnon continuum in the case where the mechanism used to time-modulate the DMI (e.g. electric fields or strain) produce an unintended modulation of the exchange parameter $J_\perp$ (see Fig.~\ref{fig:fig5}).

\subsection{Edge states} \label{sec:sec4B}

\begin{figure}
    \centering
    \includegraphics{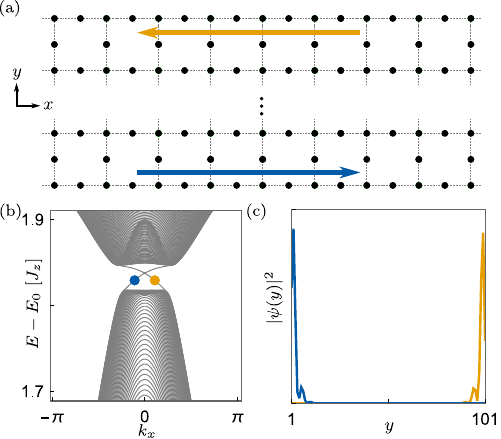}
    \caption{(a) Effective Lieb lattice with open boundary conditions in the $y$ direction and periodic boundary conditions in the $x$ direction. (b) Spectrum of the system with open boundary conditions in the $y$ direction. The (normalized) wavefunctions associated with the $k_x$ points indicated with circles are shown in (c). 
    The parameters used in this plot are $J_\perp=J_z/5$, $B =2 J_z$, $\omega = 2.5 J_z$, $d=J_z/15$.}
    \label{fig:fig6}
\end{figure}

To verify the presence of edge states we work in a ribbon geometry with periodic boundary conditions in the $x$ direction and open boundary conditions in $y$ [see Fig.~\ref{fig:fig6}(a)]. We use the partial Fourier transform
\begin{equation} \label{eq:partial_FT}
    \ket{\psi_{\bm k}} = \frac{1}{\sqrt{L}} \sum_{y} e^{i k_y y} U_{k_y}^\dagger \ket{\psi_{k_x y}},
\end{equation}
where 
$\ket{\psi_{\bm k} } = \left( \ket{\bm k},\,\, \ket{\bm k,\bm e_x},\,\, \ket{\bm k, \bm e_y} \right)^T$ is our basis vector in momentum space and $U_{k_y}^\dagger=\mathrm{diag}(1,\,1,\,e^{ik_y/2})$ ensures the Fourier transform is taken with respect to the position of each site in the Lieb lattice [staying consistent with Eq.~\eqref{eq:tmbs_y}]. The Hamiltonian for the ribbon is straightforwardly obtained by using Eq.~\eqref{eq:partial_FT} in Eq.~\eqref{eq:H_eff_3} and is given in Appendix~\ref{sec:Appendix_OBC} [Eq.~\eqref{eq:H_obc}]. In the ribbon geometry considered here, the Lieb lattice terminates in a row containing single-spin-flip states $\ket{n_x\hat{x}+L\hat{y}}$ and two adjacent spin flips in the $x$-direction $\ket{n_x\hat{x}+L\hat{y},\hat{y}}$ [see Fig.~\ref{fig:fig6}(a)]. The state $\ket{n_x\hat{x}+L\hat{x},\hat{y}}$ does not exist because it would correspond to flipping the spins at positions $(n_x,L)$ and $(n_x,L+1)$, which is outside the system boundary.

Figure~\ref{fig:fig6}(b) shows the spectrum of the system with the ribbon geometry for $B=2J_z$, $J_\perp=J_z/5$, $d=J_z/15$, $\omega = 2.5J_z$. Although the bulk spectrum is fully gapped, the ribbon supports counter-propagating edge modes that traverse the bulk gap between the lowest and middle band (Fig.~\ref{fig:fig6}). These edge modes are protected against perturbations that do not close the quasi-energy gap~\cite{zhang2013topological} which, in our system, is given by the amplitude of the modulation of the DMI $d$. 

In Floquet topological systems, anomalous edge modes traversing energy gaps between bands with trivial Chern numbers can appear, differentiating the topological bulk-boundary correspondence of systems in equilibrium \cite{rudner2020band,rudner2013anomalous,kitagawa2010topological}. These states may appear in systems where the minimum of the lowest-energy band and the maximum of the highest-energy band touch resulting in a band closure and reopening (this is allowed in Floquet systems because the quasienergy spectrum is periodic in energy) \cite{rudner2013anomalous}. In our system and in the regime under consideration: $J_\perp \ll J_z$, the frequency $\omega$ [of order $J_z$, cf. Eq.~\eqref{eq:lower_bound}] is larger than the bandwidths [of order $J_\perp$, cf. Eq.~\eqref{eq:H_eff_3} and~\eqref{eq:Delta}] and the aforementioned gap closing and reopening will not occur. Hence, there will be no anomalous edge modes in the regime under consideration. 

\subsection{Controlling chirality} \label{sec:sec4C}

We now consider the case in which the DMI can be independently time-modulated for spins that are nearest neighbors along the $x$ and $y$ directions, $D_x(t)\neq D_y(t)$ [see Eq.~\eqref{eq:H_DM}]. We focus on the case where the amplitude and frequency of the drive are the same but there is a phase difference $\varphi$ between them:
\begin{align}
    D_x(t) &= D_0 + 2d \cos(\omega t +\varphi), \label{eq:Dx}\\
    D_y(t) &= D_0 + 2d \cos(\omega t). \label{eq:Dy}
\end{align}
Going through the same procedure used to obtain Eq.~\eqref{eq:H_eff_3}, the effective Hamiltonian in the rotating frame is given by 
\begin{equation} \label{eq:H_eff_4}
    \tilde H_{\bm k}' = \left( \begin{array}{ccc}
         E_{\bm k}-\omega & -i e^{-i\varphi}ds_x  & d s_y\\
         i e^{i\varphi} d s_x & \tilde E_{\bm k}^x-2\omega & g_{\bm k}' \\
         d s_y & (\tilde g_{\bm k}')^* & \tilde E_{\bm k}^y-2\omega
    \end{array}\right),
\end{equation}
where 
\begin{equation}
    \tilde g'_{\bm k} = -2\frac{J_\perp^2}{J_z}c_xc_y-\frac{i}{2}\frac{d^2}{B+J_z-\omega}e^{i\varphi}s_xs_y.
\end{equation}

\begin{figure}
    \centering
    \includegraphics{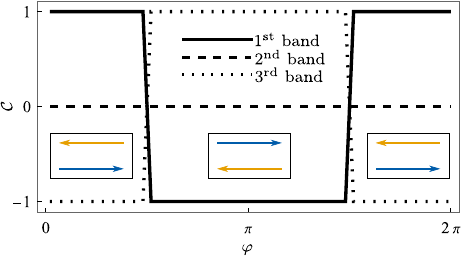}
    \caption{Chern numbers as a function of the relative phase $\varphi$ between the time-modulation of the DMI between neighboring spins in the $x$ and $y$ directions [see Eqs.~\eqref{eq:Dx} and~\eqref{eq:Dy}]. The inset show the direction of the topological edge states. The parameters used for this plot are $J_\perp=J_z/5$, $B = 2J_z$, $\omega = 2.5J_z$, $d=J_z/15$.}
    \label{fig:fig7}
\end{figure}

The phase $\varphi$ can then be used to tune the topological phase and hence the propagation direction of the magnon edge states. Figure~\ref{fig:fig7} shows the Chern numbers for the three bands of the Hamiltonian~\eqref{eq:H_eff_4} as a function of $\varphi$ and the insets show the chirality of the edge states in the different topological phases. The Chern numbers are computed using the method by Fukui \textit{et al.}, as in the previous sections. The phase transition caused by the variation of $\varphi$ can be understood by comparing the Hamiltonians obtained for $\varphi$ and $\varphi + \pi$. These Hamiltonians are related by complex conjugation: $\tilde H_{\bm k}'(\varphi)=[\tilde H_{\bm k}'(\varphi+\pi)]^*$ which means that they are related by an effective time-reversal (TR) operation. This operation consists of the real TR and a rotation of the spin operators by $\pi$ about the $x$ or $y$ axis, as shown by Mook \textit{et al.}~\cite{mook2023magnons}. As a consequence of this relation, the Berry curvatures of the eigenstates of $\tilde H_{\bm k}'(\varphi)$ and $\tilde H_{\bm k}'(\varphi+\pi)$ have opposite sign (see Appendix~\ref{app:effective_TRS}) and therefore their corresponding Chern numbers also have opposite sign (Fig.~\ref{fig:fig7}). This sign change of the Chern number results in the reversal of the chirality of the edge modes~\cite{mook2014edge}. 

\subsection{Potential experimental realization} \label{sec:sec4D}

By proposing the time modulation of the DMI, we extend the range of materials in which magnonic topological edge modes having single-magnon and two-magnon character can be found by removing the necessity of exchange interactions beyond nearest neighbors~\cite{mook2023magnons}. Here, we have used a simple model to illustrate how driving the DMI may induce a topological phase transition. The idea of driving the DMI to induce topological phase transitions may be applied to other lattices and different spin $S$. Systems with $S>1/2$ may allow for the existence of a different type of bound state, the \textit{single-ion bound state}, in the presence of single-ion anisotropy~\cite{mook2023magnons,bai2021hybridized}. 
 
Van der Waals magnets~\cite{burch2018magnetism,Park2016opportunities,ManasValero2025Fundamentals} are one class of two-dimensional materials that may be suitable for the implementation of the technique proposed here. Easy-axis anisotropy (Ising and/or single-ion) has been found, for example, in $\mathrm{Cr}_2\mathrm{Ge}_2\mathrm{Te}_6$~\cite{gong2017discovery} and $\mathrm{CrI}_3$~\cite{lado2017origin,huang2017layer}. Another set of interesting materials includes Janus monolayers~\cite{lu2017janus,zhang2017janus} which may have Ising or single-ion anisotropy and where inversion symmetry is inherently broken allowing for the DMI considered here~\cite{liang2020very,yuan2020intrinsic,xu2020topological}. Both van der Waals magnets and Janus monolayers~\cite{gong2017discovery,lado2017origin,xu2020topological,yuan2020intrinsic,liang2020very} have exchange couplings on the order of meV and therefore the frequency $\omega$ [which is on the order of $B+J_z$, cf. Eqs.~\eqref{eq:Delta} and~\eqref{eq:inequalities_Delta}] required for resonant driving lies in the THz regime.
One strategy to drive the DMI is therefore the use of a THz-frequency electric field. In van der Waals materials, it has been predicted that the DMI amplitude depends linearly on an externally applied static electric field and the strength of the DMI can be changed by up to $\sim 1$ meV using field amplitudes of $E\sim 1$ V/nm~\cite{liu2018analysis}. An oscillating electric field may then be used to drive the DMI [Eq.~\eqref{eq:Dt}] if the drive adiabatically modulates the magnetic system without generating spurious excitations. The THz radiation will also generate a direct magnetic coupling, which could resonantly couple the single-magnon state to the TMBSs. This coupling, given by $\braket{\bm k| \mu \bm B_\perp \cdot \bm S_\perp  |\bm k, \bm e_i} = \mu |\bm B_\perp| \cos (k_i/2)$ ($\mu$ is the magnetic moment of the magnetic atom and $\bm B_\perp$, $\bm S_\perp$ are the in-plane components of the magnetic field and spin operator) in the rotating frame, will not affect the topological phase of the system when its strength is small compared to the quasienergy gap of order $d$: $\mu |\bm B_\perp| \ll d$.  
The amplitude of the DMI can also be changed by up to $\sim 1$ meV by varying the polarization of ferroelectric layers in van der Waals heterostructures~\cite{chen2024dzyaloshinskii} or by applying strain~\cite{cui2020strain}. Modulation of the polarization in ferroelectrics~\cite{miyamoto2013ultrafast,mankowsky2017ultrafast,wang2025ultrafast} and strain~\cite{luo2023ultrafast,gollwitzer2025picosecond} in the THz regime has been achieved experimentally and can serve as additional strategies to drive the DMI. 
Furthermore, since the DMI depends on the same orbital overlaps as Heisenberg exchange, lattice vibrations that modulate the Heisenberg exchange interaction in the THz regime~[82,83] are also expected to modulate the DMI in this frequency range (spurious resonances that may occur when modulating the other exchange parameters can be prevented by changing the value of an external magnetic field, see Fig.~5).
We note that, although the materials mentioned above ($\mathrm{Cr}_2\mathrm{Ge}_2\mathrm{Te}_6$, $\mathrm{CrI}_3$, Janus monolayers of manganese dichalcogenides, $\mathrm{VSeTe}$ and $\mathrm{Cr(I,X)}_3$) have easy-axis anisotropy (Ising and/or single-ion), the condition $J_\perp \ll J_z$ is not satisfied in any of them and therefore this list does not contain a direct candidate material. However, the discovery of new van der Waals magnets is an active field of research~\cite{burch2018magnetism} which may lead to materials with stronger anisotropy to satisfy the condition $J_\perp \ll J_z$.

A more direct implementation could be realized using Rydberg atoms which can be used to simulate magnetic systems~\cite{Browaeys2020Manybody}. By applying a sequence of global and local pulses, it is possible to engineer the XXZ Hamiltonian with the DMI interactions of our model. The strength of the magnetic interactions depends on the time interval between the applied pulses which can be varied to generate the time-modulation of the DMI~\cite{kuji2025proposal}. Systems of cold atoms, also used to simulate magnetic systems are an additional alternative to simulate the physics proposed in this paper. In these systems, in addition to XXZ anisotropy~\cite{duan2003controlling,Trotzky2008Time}, a synthetic spin-orbit interaction can be engineered resulting in an effective spin Hamiltonian with DMI~\cite{gong2015dzyaloshinskii,radic2012exotic,cole2012bose}.   

\section{Conclusion}\label{sec:sec5}

We have shown that a periodic modulation of the DMI can be used as a strategy to produce robust topological edge modes in quantum magnets. We considered the case of a square lattice of spins with ferromagnetic longitudinal interactions, antiferromagnetic transverse interactions, and DMI. In the case of a static DMI, the magnonic system is topologically non-trivial in the sense that the bands have nonzero Chern numbers. However, robust edge modes are absent due to the lack of an energy gap between the non-trivial bands in the regime under consideration: $J_z\gg J,D_0$. We have further shown that, by driving the DMI, a band inversion between the single-magnon band and the TMBSs is produced with an energy gap that allows for the existence of robust edge states. Furthermore, we have shown that the chirality of the edge states can be controlled by driving the DMI for neighbors in the $x$ and $y$ direction with a relative phase. 

\begin{acknowledgments}
We acknowledge support from Fonds de Recherche du Qu\'ebec--Nature et Technologies (FRQNT). TP-B and WAC acknowledge support from Natural Sciences and Engineering Research Council of Canada (NSERC) and Institut Transdisciplinaire d'Information Quantique (INTRIQ).  TP-B acknowledges the hospitality of the Institut de Ciencies Fotoniques (Barcelona) where some of the work took place.
\end{acknowledgments}

\appendix

\section{Exact diagonalization of the static Hamiltonian} \label{sec:exact_diagonalization}

Figure~\ref{fig:fig1}(b) shows the spectrum of the static Hamiltonian $H_\mathrm{S}$ in the subspaces of one, two, and three spin flips. $H_\mathrm{S}$ preserves the number of spin flips and it can be diagonalized in subspaces with a definite number of spin flips. For one spin flip, there is only the single-magnon band given by Eq.~\eqref{eq:E_k}. 

To obtain the spectrum of the two-magnon states, we first need to choose a basis. Since we have translation symmetry we use the two-magnon states $\ket{\bm k,\bm r}$ defined in Eq.~\eqref{eq:tm_def}. The matrix elements are obtained by applying $H_\mathrm{S}$ to $\ket{\bm k,\bm r}$:
\begin{equation} \label{eq:H_S_on_kr}
\begin{split}
    H_\mathrm{S} & \ket{\bm k,\bm r} = \bigg(2B+4J_z -  J_z \sum_{i=x,y}\delta_{\bm r,\bm e_i} \bigg) \ket{\bm k,\bm r} \\& +J_\perp\sum_{\substack{i=x,y\\\sigma=\pm}} \cos\frac{k_i}{2} \ket{\bm k,\bm r + \sigma \bm e_i}.
\end{split}
\end{equation}
The state $\ket{\bm k,\bm r}$ is a superposition of the states with two spin flips separated by $\bm r=n_x\bm e_x + n_y \bm e_y$ [cf. Eq.~\eqref{eq:tm_def}] (where $n_x,n_y \geq 0$ are defined modulo $L$). 
For spin $1/2$, $n_x,n_y=0$ is not an option as this corresponds to applying the spin lowering operator twice on the same lattice site. 
Furthermore, from the remaining $N-1$ vectors satisfying $(n_x,n_y)\neq(0,0)$, the states $\ket{\bm k,\bm r}$ and $\ket{\bm k, L \bm e_x + L \bm e_y-\bm r}$ are the same [cf. Eq.~\eqref{eq:tm_def}]. 
For odd $L$, this means that each of the $N$ wavevectors $\bm k$ has $(N-1)/2$ options for $\bm r$ so that the size of the Hilbert space of the two-magnon states is $N(N-1)/2$, the number of ways in which two spins can be flipped. 
For even $L$, the vectors $\bm r \in \{ L/2\bm e_x,L/2\bm e_y,L/2 \bm e_x + L/2 \bm e_y \}$ are the same as $L \bm e_x + L \bm e_y -\bm r$ and for these vectors there are only $N/2$ distinct wavevectors $\bm k$ that give distinct $\ket{\bm k,\bm r}$. Hence, for even $L$, there are $3$ vectors $\bm r$ that have $N/2$ options for $\bm k$ and $(N-4)/2$ vectors $\bm r$ that have $N$ choices of $\bm k$, so that the size of the Hilber space is again $N(N-1)/2$. Figure~\ref{fig:fig1}(b) shows the spectrum for $L=7$.

To obtain the spectrum of the three-magnon states we use the basis
\begin{equation} \label{eq:3magnon_state}
\begin{split}
  &\ket{\bm k,\bm r_1,\bm r_2} =\\& \hspace{13mm} \frac{1}{\sqrt{N}} \sum_{\bm r} e^{i\bm k \cdot \left(\bm r + \frac{\bm r_1 + \bm r_2}{3}\right)}\ket{\bm r,\bm r+\bm r_1,\bm r+ \bm r_2},
\end{split}
\end{equation}
where
\begin{equation}
    \ket{\bm r,\bm r',\bm r''}\equiv S_{\bm r}^- S_{\bm r'}^- S_{\bm r''}^- \ket{0}.
\end{equation}
The matrix elements can be obtained by applying $H_\mathrm{S}$ on Eq.~\eqref{eq:3magnon_state}:
\begin{widetext}
\begin{equation} \label{eq:H_S_on_3mag}
\begin{split}
   H_\mathrm{S} \ket{\bm k,\bm r_1,\bm r_2} =
   \bigg[3 B &+ 6J_z - J_z\sum_i \big( \delta_{\bm r_1,\bm e_i} + \delta_{\bm r_2,\bm e_i} 
    + \delta_{\bm r_1-\bm r_2,\bm e_i} \big) \bigg]\ket{\bm k,\bm r_1,\bm r_2} 
    \\&+ J_\perp\sum_{\substack{i=x,y\\\sigma=\pm}} e^{-i \sigma k_i/3} \big( \ket{\bm k,\bm r_1 -\sigma \bm e_i ,\bm r_2-\sigma \bm e_i} + \ket{\bm k,\bm r_1 +\sigma \bm e_i ,\bm r_2} + \ket{\bm k,\bm r_1 ,\bm r_2+\sigma \bm e_i} \big).
\end{split}
\end{equation}    
\end{widetext}
The states $\ket{\bm k,\bm r_1,\bm r_2}$, $\ket{\bm k, -\bm r_1,\bm r_2-\bm r_1}$, and $\ket{\bm k, -\bm r_2, \bm r_1-\bm r_2}$ are the same [cf. Eq.~\eqref{eq:3magnon_state}]. Hence, the size of the Hilbert subspace for each wavevector $\bm k$ is $(N-1)(N-2)/6$. 

\section{Effective Hamiltonians} \label{sec:Appendix_effective_Hamiltonian}

\subsection{Perturbation theory}

For a time-independent Hamiltonian given by
\begin{equation}
    H = H_0 + V, \quad H_0 \ket{\alpha} = \varepsilon_\alpha \ket{\alpha},
\end{equation}
where the eigenvalues $\varepsilon_\alpha$ and eigenstates $\ket{\alpha}$ of $H_0$ are known, an effective Hamiltonian describing the low-energy subspace can be obtained using perturbation theory. The matrix elements of this effective Hamiltonian, up to second-order corrections in $V$, are given by (see, for example, complement BI of Ref.~\cite{cohen1998atom})
\begin{equation} \label{eq:H_eff_0}
\begin{split}
    H^\mathrm{eff}_{\alpha\beta} = &(H_0+V)_{\alpha\beta}  \\&+ \frac{1}{2}\sum_{\gamma}V_{\alpha\gamma}V_{\gamma\beta}\left(\frac{1}{\varepsilon_\alpha-\varepsilon_\gamma}+\frac{1}{\varepsilon_\beta-\varepsilon_\gamma}\right),
\end{split}
\end{equation}
where the indices $\alpha\beta$ label states in the low-energy subspace and $\gamma$ are higher-energy states out of the low-energy subspace of interest. 

\subsection{Effective Hamiltonian in the static case} \label{sec:app_Eff_H_hs}

To obtain the Hamiltonians~\eqref{eq:H_eff_1} and~\eqref{eq:H_eff_2} corresponding, respectively, to the case of vanishing DMI ($D=0,d=0$) and the case of time-independent DMI ($d=0$). We take the Ising part of $H(t)$ as the unperturbed Hamiltonian and the transverse exchange interactions and the DMI as perturbations. In the the case of a static DMI the Hamiltonian $H(t)$ becomes time independent:
\begin{subequations} 
\begin{equation} \label{eq:perturbation_splitting}
    H(t) \rightarrow H = H_0 + V_1 + V_2,
\end{equation}
where
\begin{align}
    H_0 &= -J_z\sum_{\braket{\bm r,\bm r'}}S_{\bm r}^z S_{\bm r'}^z-B\sum_{\bm r} S_{\bm r}^z,\\
    V_1 &= J_\perp \sum_{\braket{\bm r,\bm r'}}\left(S_{\bm r}^x S^x_{\bm r'} + S_{\bm r}^y S^y_{\bm r'} \right),\\
    V_2 &= D_0 \sum_{\bm r} \big[ (\bm S_{\bm r} \times \bm S_{\bm r+\bm e_x} )_y + \bm  (\bm S_{\bm r} \times \bm S_{\bm r+\bm e_y} )_x \big].
\end{align}
\end{subequations}
The treatment of $V_1,V_2$ as perturbations is justified in the limit 
\begin{equation}
    J_z\gg J,D_0,
\end{equation}
as the energy difference between the low-energy subspace \{$\ket{\bm k}$, $\ket{\bm k,\bm e_x}$, $\ket{\bm k,\bm e_y}$\} and the (higher-energy) states coupled to it by $V_1,V_2$ is at least $J_z$. To obtain Eqs.~\eqref{eq:H_eff_1} and~\eqref{eq:H_eff_2} we use Eq.~\eqref{eq:H_eff_0} with $V=V_1$ and $V=V_1+V_2$, respectively.

\subsection{Effective Hamiltonian with time-dependent DMI} \label{sec:app_Eff_H_ht}

To obtain the effective Hamiltonian in the case of the time-modulated DMI, we go into a rotating frame given by $\ket{\psi(t)} = U(t) \ket{\tilde \psi(t)}$ where
\begin{equation}
    U(t) = e^{-i\omega S_\mathrm{sf} t},\quad S_\mathrm{sf} = \frac{N}{2}-\sum_{\bm r}S_{\bm r}^z,
\end{equation}
where $S_\mathrm{sf}$ counts the number of spin flips. 
The Hamiltonian dictating the time-evolution of the states $\ket{\tilde \psi(t)}$ in the rotating frame is given by 
\begin{equation}
\begin{split}
    \tilde H(t) &= U^\dagger(t)H(t)U(t)-\omega S_\mathrm{sf} \\&= H_\mathrm{S} - \omega S_\mathrm{sf} + U^\dagger(t) H_\mathrm{DM}(t) U(t),
\end{split}
\end{equation}
where $H(t)=H_\mathrm{S}+H_\mathrm{DM}(t)$ is given by Eq.~\eqref{eq:Ht}. Since $H_\mathrm{S}$ preserves the number of spin flips we have $U^\dagger(t)H_\mathrm{S}U(t)=H_\mathrm{S}$. The DMI couple states whose number of spin flips differ by one. Defining $P_n$ as the projection operator onto the subspace with $n$ spin flips, the equation above can be written as
\begin{widetext}
\begin{equation} \label{eq:rotating_dmi}
\begin{split}
    U^\dagger(t) H_\mathrm{DM}(t) U(t) &= e^{i\omega S_\mathrm{sf}} \left(D_0 + d  e^{i\omega t} + d e^{-i\omega t}\right) \sum_{\bm r} \big[ (\bm S_{\bm r} \times \bm S_{\bm r+\bm e_x} )_y + \bm  (\bm S_{\bm r} \times \bm S_{\bm r+\bm e_y} )_x \big] e^{-i\omega S_\mathrm{sf}} \\ &= \sum_n \left\{ P_n \left[
    \left(e^{-i\omega t} D_0 + d+ d e^{-2i\omega t}\right) \sum_{\bm r} \big[ (\bm S_{\bm r} \times \bm S_{\bm r+\bm e_x} )_y + \bm  (\bm S_{\bm r} \times \bm S_{\bm r+\bm e_y} )_x \big]  \right] P_{n+1} + \mathrm{h.c.}\right\}. 
\end{split}
\end{equation}
\end{widetext}
As explained in the main text we focus on the regime where the single magnon band is resonantly coupled to the TMBSs and the TMBSs is off-resonant from the three-magnon states [see Fig.~\ref{fig:fig1}(b)]. Neglecting counter-rotating terms, we have
\begin{equation}
\begin{split}
    U^\dagger(t) H_\mathrm{DM}(t) U(t) &\approx \\
    d \sum_{\bm r} \big[ (\bm S_{\bm r} &\times \bm S_{\bm r+\bm e_x} )_y + \bm  (\bm S_{\bm r} \times \bm S_{\bm r+\bm e_y} )_x \big]
\end{split} 
\end{equation}
The Hamiltonian in the rotating frame then is the same as the Hamiltonian with static DMI, now with strength $d$ instead of $D_0$, and where the $n$-magnon bands are shifted $n\omega$ downwards in energy, bringing bands with different number of magnons closer to each other. We then apply the same perturbative approach as we did to obtain the effective Hamiltonian for the static DMI to obtain Eq.~\eqref{eq:H_eff_3}. 
In the case where a phase between the DMI in the $x$ and $y$ directions [$D_x(t)\neq D_y(t)$] is present the DMI in the rotating-wave approximation is given by
\begin{equation}
\begin{split}
    U^\dagger(t) & H_\mathrm{DM}' U(t) \approx d \sum_{\bm r }   (\bm S_{\bm r}\times S_{\bm r+\bm e_y})_x \\& +d \sum_{\bm r n} \left[e^{i\varphi} P_{n}  (\bm S_{\bm r}\times \bm S_{\bm r+\bm e_x})_y P_{n+1} + \mathrm{h.c.} \right] , 
\end{split}
\end{equation}
which can be seen from adding the corresponding phase in Eq.~\eqref{eq:rotating_dmi}.

\section{Chern numbers and effective time-reversal symmetry} \label{sec:Appendix_time_reversal}

\subsection{Berry curvature}

The topology of a two-dimensional system without symmetry restrictions, other than translation symmetry, is characterized by the Chern of its energy bands~\cite{chiu2016classification}. Let us consider the Hamiltonian $\hat H$ (in this appendix we follow Ref.~\cite{chiu2016classification} and use the hat notation $\hat O$ to explicitly distinguish operators/operator vectors from vectors and matrices whose elements are complex numbers) which is diagonal in the momentum basis:
\begin{equation} \label{eq:hat_H}
    \hat{\mathcal{H}} = \sum_{\bm k} \hat \psi^\dagger_{\bm k} \mathcal H_{\bm k} \hat \psi_{\bm k} = \sum_{\bm k} \hat \psi^\dagger_{\bm k\alpha} (\mathcal H_{\bm k})_{\alpha\beta} \hat \psi_{\bm k \beta},
\end{equation}
where $\hat \psi_{\bm k}^\dagger$ is a vector of creation operators with components $\hat \psi^\dagger_{\bm k \alpha}$ with $\alpha=1,\dots,M$, where $M$ is the number of internal degrees of freedom (e.g. number of atoms in the unit cell) and $\mathcal H_{\bm k}$ is the ($M\times M$) matrix characterizing the Hamiltonian. In the equation above and for the rest of this appendix we use Einstein summation notation for Greek indices. 

The eigenstates of $\hat{\mathcal{H}}$ may be obtained by computing the eigenstates of $\mathcal H_{\bm k}$. The eigenstate from the $n^\mathrm{th}$ band are obtained by solving the eigenvalue equation
\begin{equation} \label{eq:hk_phi}
    \mathcal H_{\bm k} \phi_{n\bm k} = \varepsilon_n \phi_{n\bm k}.
\end{equation}
Then, the Chern number $\mathcal{C}_n$ for this band is obtained by integrating the $z$ component of the Berry curvature over the Brillouin zone (BZ)~\cite{asboth2016short,xiao2010berry}
\begin{equation} \label{eq:Chern_number}
    \mathcal{C}_n = \frac{1}{2\pi} \int_\mathrm{BZ} d^2\bm k \,\mathcal{B}_{n\bm k,}^z, 
\end{equation}
where 
\begin{equation} \label{eq:bc_k}
    \mathcal{B}_{n\bm k}^k = -\epsilon_{ijk} \, \mathrm{Im}\left[ \partial_i \phi_{n\bm k \alpha}^*\partial_j \phi_{n\bm k \alpha} \right]
\end{equation}
is the $k^\mathrm{th}$ component of the Berry curvature of the $n^\mathrm{th}$ band at momentum $\bm k$, $\phi_{n\bm k \alpha}$ is the $\alpha^\mathrm{th}$ component of the vector $\phi_{n\bm k}$, $\partial_i = \partial/\partial k_i$ is the derivative with respect to the $i^\mathrm{th}$ component of the momentum, and $\epsilon_{ijk}$ is the Levi-Civita symbol in three dimensions (the indices $i,j,k$ take the values $x,y,z$). 

\subsection{Effective TRS} \label{app:effective_TRS}

The system described by the Hamiltonian $\hat{\mathcal H}$ [Eq.~\eqref{eq:hat_H}] has an effective time-reversal symmetry if there is a unitary matrix $U$ acting on the internal degrees of freedom and an orthonormal $3\times 3$ matrix $O$ such that 
\begin{equation} \label{eq:TRS_transformation}
    U^\dagger \mathcal H_{\bm k'} U = \mathcal H_{\bm k}^*, \quad \bm k'=O\bm k, 
\end{equation}
is satisfied. If this symmetry is present, the system is topologically trivial in the sense that its Chern numbers vanish. 

To show this, we note that when Eq.~\eqref{eq:hk_phi} is satisfied then the equation
\begin{equation}
    \mathcal H_{\bm k}^* \phi_{n \bm k }^*= \varepsilon_n \phi_{n\bm k }^*,
\end{equation}
is also satisfied. By acting with $U$ on the left and using Eq.~\eqref{eq:TRS_transformation} we obtain
\begin{equation}
    U\mathcal H_{\bm k}^* \phi_{n\bm k}^*  =  \mathcal H_{\bm k'} U \phi_{n \bm k}^* =  \varepsilon_n U \phi_{n \bm k}^* ,
\end{equation}
where we used $UU^\dagger = 1$ as $U$ is a unitary transformation. Therefore if $\phi_{n\bm k}$ is an eigenstate of $\mathcal H_{\bm k}$ then $U\phi^*_{n \bm k}$ is an eigenstate of $\mathcal H_{\bm k'}$ with the same energy $\varepsilon_n$ given that the effective TRS symmetry~\eqref{eq:TRS_transformation} is present.

Let us now compare the Berry curvature at $\bm k$ and $\bm k'$ for a given band $n$. Using Eq.~\eqref{eq:bc_k} and the fact that $U\phi_{\bm k n}$ is the $n^\mathrm{th}$ eigenstate of $h_{\bm k'}$ we have
\begin{equation}
\begin{split}
    \mathcal{B}_{n \bm k'}^z &= - \epsilon_{ijk}\, \mathrm{Im} \left[ \partial_i (U_{\alpha\beta }\phi_{n\bm k\beta }^*)^* \partial_j (U_{\alpha \beta}\phi_{n\bm k\beta }^*)\right] \\
    & = - \epsilon_{ijk}\, \mathrm{Im} \left[ \partial_i (U_{\beta \alpha}^\dagger\phi_{n\bm k\beta }) \partial_j (U_{\alpha \beta}\phi_{n\bm k\beta }^*)\right] \\ 
    & = - \epsilon_{ijk}\, \mathrm{Im} \left[ \partial_i\phi_{n\bm k\alpha } \partial_j \phi_{n\bm k\alpha }^*\right] \\ 
    & = - \epsilon_{jik}\, \mathrm{Im} \left[ \partial_j\phi_{n\bm k\alpha } \partial_i \phi_{n\bm k\alpha }^*\right],
\end{split}
\end{equation}
where in the third line we used $U^\dagger_{\beta\alpha} U_{\alpha\beta}=\delta_{\alpha\beta}$. Comparing the equation above with Eq.~\eqref{eq:bc_k} and noting that $\epsilon_{jik}=-\epsilon_{ijk}$ we obtain
\begin{equation}
    \mathcal{B}_{n\bm k'}^k=-\mathcal{B}_{n\bm k}^k.
\end{equation}
Therefore, the contributions to the integral~\eqref{eq:Chern_number} at momenta $\bm k$ and $\bm k'$ cancel each other and the Chern numbers vanish. 

\section{Three-magnon bound states} \label{sec:3mbs_perturbation}

In this section we obtain the lowest energy of the three-magnon states in order to find the detuning $\Delta$ between the TMBSs and the three-magnon states when the DMI is driven. We use first-order perturbation theory where the unperturbed Hamiltonian $H_0$ and the perturbation $V_1$ are given by Eq.~\eqref{eq:perturbation_splitting}. In the limit $J_z\gg J_\perp$ the three-magnon spectrum splits into three branches: three-magnon bound states, one TMBSs and a single magnon and 3 single-magnon states. The three-magnon bound states have the lowest energy and are given by the Fourier transform of three adjacent spin flips. There are six states with three neighboring spin flips with unperturbed energy $3B+4J_z$, these are 
\begin{equation*}
    \begin{tikzpicture}
        \protect\filldraw[black] (0,0) circle (2pt); 
        \protect\filldraw[black] (10pt,0) circle (2pt); 
        \protect\filldraw[black] (0,10pt) circle (2pt); 
        \protect \draw (0,0) -- (10pt,0);
        \protect \draw (0,0) -- (0,10pt);
    \end{tikzpicture} 
    \qquad  
    \begin{tikzpicture}
        \protect\filldraw[black] (0,0) circle (2pt); 
        \protect\filldraw[black] (10pt,0) circle (2pt); 
        \protect\filldraw[black] (10pt,10pt) circle (2pt); 
        \protect \draw (0,0) -- (10pt,0);
        \protect \draw (10pt,0) -- (10pt,10pt);
    \end{tikzpicture} 
    \qquad 
    \begin{tikzpicture}
        \protect\filldraw[black] (0,0) circle (2pt); 
        \protect\filldraw[black] (0,10pt) circle (2pt); 
        \protect\filldraw[black] (-10pt,10pt) circle (2pt); 
        \protect \draw (0,0) -- (0,10pt);
        \protect \draw (0,10pt) -- (-10pt,10pt);
    \end{tikzpicture} 
    \qquad 
    \begin{tikzpicture}
        \protect\filldraw[black] (0,0) circle (2pt); 
        \protect\filldraw[black] (0,10pt) circle (2pt); 
        \protect\filldraw[black] (10pt,10pt) circle (2pt); 
        \protect \draw (0,0) -- (0,10pt);
        \protect \draw (0,10pt) -- (10pt,10pt);
    \end{tikzpicture} 
    \qquad 
    \begin{tikzpicture}
        \protect\filldraw[black] (0,0) circle (2pt); 
        \protect\filldraw[black] (10pt,0) circle (2pt); 
        \protect\filldraw[black] (20pt,0) circle (2pt); 
        \protect \draw (0,0) -- (10pt,0);
        \protect \draw (10pt,0) -- (20pt,0);
    \end{tikzpicture} 
    \qquad
    \begin{tikzpicture}
        \protect\filldraw[black] (0,0) circle (2pt); 
        \protect\filldraw[black] (0,10pt) circle (2pt); 
        \protect\filldraw[black] (0,20pt) circle (2pt); 
        \protect \draw (0,0) -- (0,10pt);
        \protect \draw (0,10pt) -- (0,20pt);
    \end{tikzpicture}
\end{equation*}
where each black dot represents a spin flip. 
The first four states are coupled to first order in $V_1$ since only the hopping of one spin flip (represented by the black dot) is required to go from one state to another. In the subspace of these four state, which in momentum space are 
\begin{equation} \label{eq:4_relevant_3mbs}
    \left(\begin{matrix}
       \ket{\bm k,\hat{x},\hat{y}} & \ket{\bm k,\hat{x},\hat{x}+\hat{y}}& \ket{\bm k,-\hat{x},-\hat{y}}  & \ket{\bm k,\hat{y},\hat{x}+\hat{y}}
    \end{matrix}\right),
\end{equation}
the perturbation $V_1$ is given by
\begin{equation}
    V_1 = \frac{J_\perp}{2}\left( \begin{matrix}
        0 & e^{ik_x/3} & 0&e^{ik_y/3} \\
        e^{-ik_x/3} & 0 & e^{ik_y/3} & 0 \\
        0& e^{-ik_y/3} & 0  & e^{-ik_x/3} \\
        e^{-ik_y/3}&0 & e^{ik_x/3}&0 \\
    \end{matrix}\right),
\end{equation}
which has eigenvalues $3B+4J_z$ (doubly degenerate), $3B+4J_z+J_\perp$ and $3B+4J_z-J_\perp$. Hence, the lowest-energy three-magnon state is (to first order in $V_1$) $3B+4J_z-J_\perp$. 

\section{Hamiltonian in the ribbon geometry} \label{sec:Appendix_OBC}
To write the Hamiltonian for the ribbon geometry we define the vector $\psi_{\bm k} = (\ket{\bm k}\, \ket{\bm k,\bm e_{\bm x} }\, \ket{\bm k,\bm e_{\bm y}})^T$. Then, the effective Hamiltonian with periodic boundary conditions can be written as 
\begin{equation} \label{eq:H_3}
    \tilde H = \sum_{\bm k}\psi^\dagger_{\bm k} \tilde H_{\bm k} \psi_{\bm k}, 
\end{equation}
where $\tilde H_{\bm k}$ is given by Eq.~\eqref{eq:H_eff_3}. By using the partial Fourier transform
\begin{equation} \label{eq:partial_FT_psi}
    \psi_{k_y y} =\frac{1}{L}\sum_{k_y} e^{ik_y y} U_{k_y} \psi_{\bm k},
\end{equation}
where $U_{ky}=\mathrm{diag}(1,1,e^{-ik_y/2})$. in Eq.~\eqref{eq:H_3} we obtain the Hamiltonian with open boundary conditions:
\begin{equation} \label{eq:H_obc}
    H_\mathrm{op} = \sum_{k_x}\sum_{y=1}^{L-1} \left[ \psi_{k_xy}^\dagger  \bm \mu  \psi_{k_xy} +  \left( \psi_{k_xy+1}^\dagger  \bm T \psi_{k_xy} + \mathrm{h.c.} \right) \right],
\end{equation}
where the matrices $\bm \mu$ and $\bm T$ are given by
\begin{equation}
    \bm \mu = \left( 
    \begin{array}{ccc}
        \mu_{11} & -ids_x & -id/2  \\
        i ds_x & \mu_{22} & \frac{d^2s_x}{2\delta}-\frac{J_\perp^2c_x}{J_z} \\
        id/2 &  \frac{d^2s_x}{2\delta} -\frac{J_\perp^2c_x}{J_z} & \mu_{33}
    \end{array}
    \right),
\end{equation}
where 
\begin{align}
    \begin{split}
    \mu_{11} &=  B+2J_z-\omega - J_\perp \cos k_x,
    \end{split}
    \\[2ex]
    \begin{split}
    \mu_{22} &= 2B + 3J_z -2\omega - \frac{J_\perp^2(1+c_x^2)}{J_z} \\&\hspace{40mm} + \frac{d^2\left( 2-c_x^2 \right)}{\delta },
    \end{split}
    \\[2ex]
    \begin{split}
    \mu_{33} &= 2B+3J_z -2\omega - \frac{J_\perp^2(1+c_x^2)}{J_z}  + \frac{d^2}{2\delta},
    \end{split}
\end{align}
where $\delta = B+J_z-\omega$ and
\begin{equation}
    \bm T = -\left( \begin{array}{ccc}
         J_\perp/2 & 0 & -Id^2/2 \\
         0& J_\perp^2/2J_z &  J_\perp^2c_x/J_z + d^2s_x/2\delta \\
         0& 0& (J_\perp^2/J_z+d^2/\delta)/4
    \end{array} \right).
\end{equation}


%

\end{document}